\documentclass[twocolumn,amsmath,amssymb,aps,longbibliography]{revtex4-2}
\usepackage{lineno,hyperref}
\usepackage{bm}
\usepackage{amsmath}
\usepackage{gensymb}
\usepackage{epstopdf}
\usepackage{booktabs}
\usepackage{float}

\usepackage{graphicx}
\usepackage{ulem}
\usepackage{xcolor}
\begin{document}

\title{Restructuring disorder: Transformation from the antiferromagnetic order in Fe$_2$VSi to the ferromagnetic state in FeRuVSi by substitution of a non-magnetic element}

\author{Shuvankar Gupta$^{1}$}
\email{guptashuvankar5@gmail.com}
\author{Sudip Chakraborty$^{1}$}
\author{Celine Barreteau$^2$}
\author{Jean-Claude Crivello$^{2,3}$}
\author{Jean-Marc Greneche$^4$}
\author{Eric Alleno$^2$}
\author{Chandan Mazumdar$^{1}$}
\email{chandan.mazumdar@saha.ac.in}

\affiliation{$^1$Condensed Matter Physics Division, Saha Institute of Nuclear Physics, 1/AF, Bidhannagar, Kolkata 700064, India}
\affiliation{$^2$Univ Paris Est Creteil, CNRS, ICMPE, UMR 7182, 2 rue H. Dunant, 94320 Thiais, France}
\affiliation{$^3$CNRS-Saint-Gobain-NIMS, IRL 3629, Laboratory for Innovative Key Materials and Structures (LINK), 1-1 Namiki, 305-0044 Tsukuba, Japan}
\affiliation{$^4$Institut des Mol\'{e}cules et Mat\'{e}riaux du Mans, IMMM, UMR CNRS 6283, Le Mans Universit\'{e}, Avenue Olivier Messiaen, Le Mans Cedex 9, 72085, France}

\date{\today}
\begin{abstract}

The delicate nature of the half-metallic ferromagnetic (HMF) property in Heusler alloys is often compromised by inherent structural disorder within the systems. Fe$_2$VSi is a prime example, where such disorder prevents the realization of the theoretically proposed HMF state as the anti-site disorder leads to the formation of two anti-parallel magnetic lattices resulting in antiferromagnetic order.
In this study, we propose an innovative and simple strategy to prevent this atomic disorder by replacing 50\% of the magnetic element Fe by a large, isoelectronic, non-magnetic element, Ru. In this way, one of the magnetic sublattices of the antiferromagnetic lattice ceases to order while ferromagnetic order is restored, an essential criterion for exhibiting HMF properties. Through various experimental measurements and theoretical calculations, we have shown that such partial replacement of Fe by Ru prevents the cross-site substitution of V/Si sites and the system regains its ferromagnetic order. Our theoretical calculations suggest that a perfect structural arrangement in Fe and Ru would have restored the HMF property in FeRuVSi. However, the local atomic disorder of Fe and Ru was found to decrease the spin polarization value. The present work sheds light on the complex interplay between structural disorder and magnetic properties in Heusler alloys and provides insights for future design strategies in the pursuit of robust half-metallic ferromagnets.

\end{abstract}
\maketitle

\section{\label{sec:Introduction}Introduction}

Half-metallic ferromagnets (HMFs) stand at the forefront of materials research due to their unique electronic properties, in particular their ability to exhibit 100\,\% spin polarization at the Fermi level~\cite{de1983new,katsnelson2008half}. As a result, HMFs have received significant attention for their potential applications in the expanding field of spintronics, where the efficient control and manipulation of electron spin is essential~\cite{felser2007spintronics}. Among the various classes of materials, Heusler alloys have emerged as one of the most promising candidates for realizing HMFs due to their versatile electronic structures and tunable magnetic properties~\cite{bombor2013half,bainsla2015spin,bainsla2015origin,gupta2022coexisting,PhysRevB.108.045137,gupta2023comncrga,PhysRevB.107.184408,gupta2024spin,kundu2017new,mondal2018ferromagnetically}.

Heusler alloys, represented by the general formula \textit{X$_2$YZ}, comprise transition metal elements, \textit{X} and \textit{Y}, a \textit{sp}-group element, \textit{Z}. These alloys typically crystallize in the \textit{L}2$_1$ structure (space group: ${Fm\bar{3}m}$, no. 225), characterized by a specific atomic arrangement where \textit{X} occupies the 8\textit{c} (0.25,0.25,0.25) position, \textit{Y} at 4\textit{b} (0.50,0.50,0.50) position, and \textit{Z} at 4\textit{a} (0,0,0) position~\cite{graf2011simple,chakraborty2023observation,chakraborty2024rare}. In the ordered structure, Heusler alloys containing a single magnetic element (\textit{X} or \textit{Y}) generally exhibit ferromagnetic ordering, while alloys with two magnetic atoms (\textit{X} and \textit{Y}) display antiferromagnetic ordering with \textit{X} and \textit{Y} magnetic moments align anti-parallel~\cite{graf2011simple,felser2015basics}. Since the Heusler alloys are made of close-by 3\textit{d}-elements having similar sizes, they are highly susceptible to atomic cross-substitution, resulting in structural disorder~\cite{gupta2022coexisting,PhysRevB.108.045137,gupta2023comncrga,gupta2024spin,PhysRevB.107.184408,PhysRevB.108.054405,PhysRevB.94.214423,chakraborty2023origin,chakraborty2023observation,smith2024effects}. In such a scenario, an antiferromagnetic ordering can also be realized in a full Heusler alloy containing a single magnetic atom if 50\% of these atoms migrate to another position and their magnetic moments aligned anti-parallel to those remaining in the predefined position.

Fe$_2$VSi serves as a paradigmatic example of this scenario. In the hypothetical ordered structure where Fe, V and Si atoms would occupy respectively the 8\textit{c}, 4\textit{b} and 4\textit{a} positions, the compound is theoretically predicted to be a HMF having a magnetic moment of  1 ${\mu_B}$~\cite{joshi2021first,abuova2022electronic}. However, in contrast to the first-principle calculations, experimental reports rather establishes Fe$_2$VSi to be an antiferromagnetic compound. Structural investigations reveal that Fe$_2$VSi crystalizes in a chemically disordered structure, in which 50\% of Fe atoms from the 8\textit{c} sites migrate to the 4\textit{a}/4\textit{b} positions, giving rise to this antiferromagnetic state~\cite{endo1995antiferromagnetism,fujii2004antiferromagnetism,fukatani2011epitaxial,nishihara2003complex,nishihara2004nmr}. The structural disorder is thus responsible for the non-adherence of the theoretically proposed HMF properties, resulting in antiferromagnetic ordering. To obtain HMF properties in such scenarios demands either the preparation of a completely ordered structure of Fe$_2$VSi, which is an extremely difficult task even in the single crystalline state~\cite{mende2021large}. The second option would be to limit or prevent the antiparallel alignment of 50\% of those Fe atoms that have migrated from the 8\textit{c} site to the 4\textit{a}/4\textit{b} sites, which would also require a crystal engineering process.

To overcome this complex problem, we have thus adopted a rather intuitive, but innovative approach, where 50\% of the Fe atoms in Fe$_2$VSi are substituted by isoelectronic but non-magnetic Ru atoms. The idea of such a replacement stems from the expectation that one of the antiferromagnetic sublattices of Fe$_2$VSi would then be non-magnetic in FeRuVSi, resulting in a ferromagnetic ground state. Since the atomic sizes of Fe and Ru are quite different, the former being a 3\textit{d}-element while the later being a 4\textit{d}-element, it is not illogical to expect that only one type of element would migrate to the 4\textit{a}/4\textit{b} position, if the migration~\cite{PhysRevB.108.054405} takes place at all.

In this study, we started with structural optimization in order to test and validate the feasibility of the above approach.
The Density Functional Theory (DFT) calculations indeed indicate a very large negative enthalpy of formation for the material, implying its plausible experimental realization. Consistent with this framework, these first calculations also predict a highly spin-polarized ground state for FeRuVSi.
Following these theoretical analyses of the ideally ordered structure, we synthesized a sample in single phase form and carried out a comprehensive study, including structural characterizations, magnetic and electrical property measurements, complemented by theoretical calculations on this novel quaternary FeRuVSi Heusler compound, taking account of its experimentally determined crystal structure.

\section{Methods}

\subsection{Experimental}

The polycrystalline FeRuVSi was synthesized via an arc melting process, employing high-purity constituent elements ($>$99.99 \,\%) under argon atmosphere. The sample underwent 5-6 melting cycles, with flipping between each melt to ensure composition homogeneity. Room temperature powder X-ray diffraction (XRD) analysis was conducted using Cu-K$\alpha$ radiation on a TTRAX-III diffractometer (Rigaku, Japan). The sample's single-phase composition and crystal structure were determined by performing a Rietveld refinement~\cite{rodriguez1993recent} using the FULLPROF software package.

Magnetic measurements were carried out in the temperature range of 3--380 K and under magnetic fields up to 70 kOe\@ using a commercial SQUID-VSM instrument (Quantum Design Inc., USA). Resistivity measurements were performed using the standard four-probe technique with a Physical Property Measurement System (Quantum Design Inc., USA).

 ${^{57}}$Fe transmission M\"{o}ssbauer spectrometry was conducted at room temperature (300 K) to check the local atomic environment of Fe atoms in the sample. Spectrum was collected using an electromagnetic transducer with a triangular velocity profile and a ${^{57}}$Co source diffused into a Rh matrix within a bath cryostat. The samples consisted of a thin powder layer containing about 5 mg Fe/cm$^2$. The hyperfine structures were modeled using a least-square fitting procedure, involving quadrupolar doublets composed of Lorentzian lines, utilizing the in-house program `MOSFIT'. Isomer shift values were reported with reference to ${\alpha}$--Fe at 300 K, and the velocity was controlled using a standard of ${\alpha}$--Fe foil.

\subsection{Computational}

The DFT method is used to evaluate the stability and fundamental properties of the materials. It is implemented in a way like that described in previous articles~\cite{gupta2022coexisting,PhysRevB.108.045137,gupta2023comncrga,PhysRevB.108.054405}. To tackle disordered structures, supercells were constructed based on the concept of special quasirandom structures (SQS) as also explained in ref.~\cite{gupta2022coexisting,PhysRevB.108.045137,gupta2023comncrga,PhysRevB.108.054405}.

\section{Results and Discussion}

\begin{table}
\caption{Calculated enthalpy of formation ($\Delta_f{H}$) for three different ordered types and one disordered structural atomic arrangement for FeRuVSi.}
\begin{tabular}{|c|c|c|c|c|c|}
\hline
 & 4\textit{a} & 4\textit{b} & 4\textit{c} & 4\textit{d} & $\Delta_f{H}$ $(kJ/mol)$ \\
\hline
Type-1 & Si & V & Ru & Fe & -51.49\\
\hline
Type-2 & Si & Ru & V & Fe & -18.89 \\
\hline
Type-3 & Si & Fe & V & Ru & -38.83\\
\hline
Disordered & Si & V & Fe:Ru & Fe:Ru & -54.29 \\
\hline
\end{tabular}\\
\label{Enthalpy}
\end{table}

\begin{figure*}[]
\centerline{\includegraphics[width=.98\textwidth]{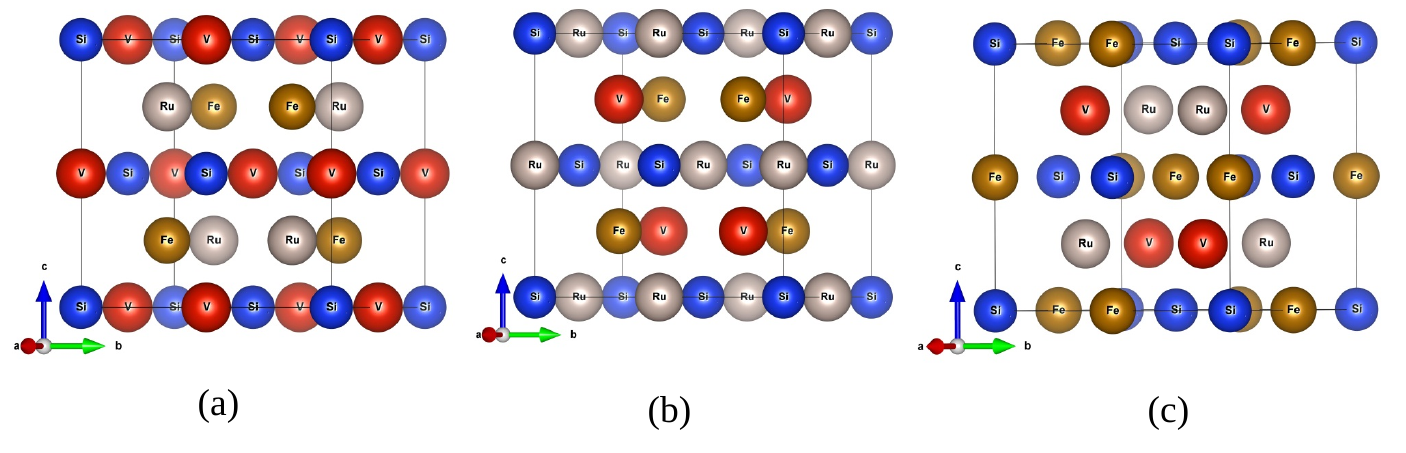}}
{\caption{Atomic arrangement of FeRuVSi for (a) Type-1 (b) Type-2 (c) Type-3 structure.}\label{Fig_1}}
\end{figure*}


\subsection{\label{sec:Electronic_Structure_Ordered} Structure optimization and electronic structure calculations}

\begin{figure*}[]
\centerline{\includegraphics[width=.98\textwidth]{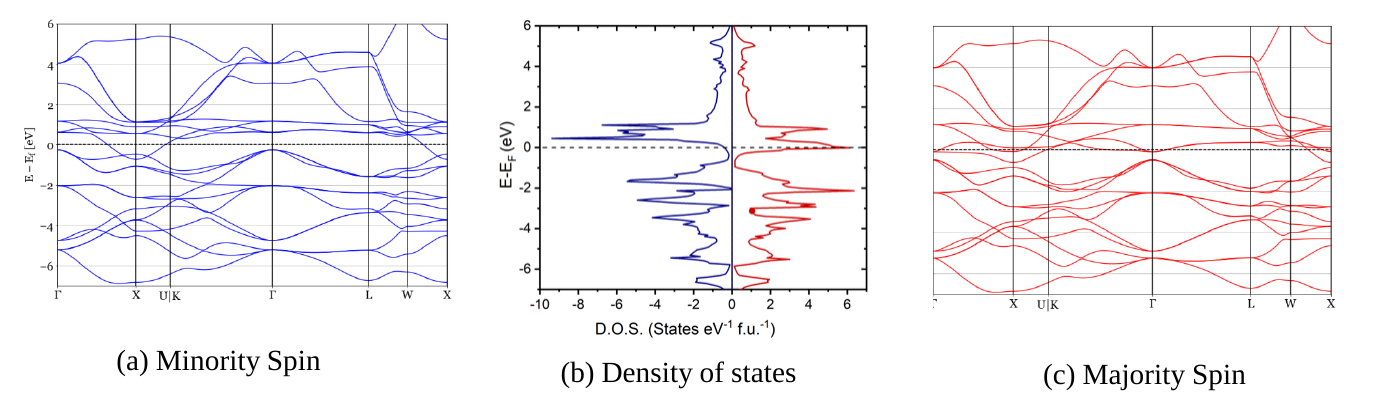}}
{\caption{Spin-polarized band structure and density of states of FeRuVSi in ordered Type-1 structure: (a) minority spin band (b) density of states, (c) majority spin band.}\label{DOS}}
\end{figure*}

The Heusler crystal structure comprises four interpenetrating face-centered-cubic (\textit{fcc}) sublattices. Notably, when one of the $X$ atoms is replaced with a different transition element denoted as $X'$ in $X_2YZ$, a quaternary Heusler alloy ($XX'YZ$) is formed. This alloy adopts a crystal structure of the LiMgPdSn-type, featuring four interpenetrating sublattices with distinct atoms~\cite{bainsla2016equiatomic}. Consequently, the $8c$ site in the $Fm\bar{3}m$ (no. 225) space group undergoes a split into $4c$ and $4d$ sites within the $F\bar{4}3m$ space group (no. 216). DFT calculations were employed to investigate various ordered atomic arrangements within the LiMgPdSn-structure type for FeRuVSi. In the quaternary $XX'YZ$ Heusler alloy, the structural configuration where the \textit{Z}-element occupies the 4\textit{a} (0,0,0) site and the three other $Y$, $X'$, and $X$ elements occupies the 4\textit{b} (0.5, 0.5, 0.5), 4\textit{c} (0.25, 0.25, 0.25), and 4\textit{d} (0.75, 0.75, 0.75) sites respectively, was initially considered. It was observed that switching atoms between the 4\textit{c} and 4\textit{d} sites resulted in energetically degenerate configurations, leading to a total of six possible ordered FeRuVSi structures. However, only three distinct structures were identifiable, as shown in Table~\ref{Enthalpy} and illustrated in Fig.~\ref{Fig_1}. These types of structure optimization calculations for quaternary Heusler alloys are very common and can be found in several reports ~\cite{gupta2022coexisting,PhysRevB.108.045137,gupta2023comncrga,PhysRevB.108.054405,gupta2023comncrga}.

Analysis of the DFT-calculated enthalpy of formation, presented in Table~\ref{Enthalpy}, revealed that the Type-1 structure (with Fe and Ru in the same planes) is more stable than the Type-2 and Type-3 structures. This preference for the Type-1 configuration is consistent with the site preferences observed in other quaternary Heusler compounds~\cite{gupta2022coexisting,PhysRevB.108.045137,gupta2023comncrga,PhysRevB.108.054405}, where the least electronegative atoms tend to occupy the 4\textit{b} site. The calculated density of states (DOS) as shown in Fig.~\ref{DOS} for the energetically most favorable Type-1 ordered structure clearly shows the half-metallic ferromagnetic nature of the compound displaying very high spin-polarization of $P=\frac{\rm{DOS}^\uparrow (E_{\rm F})- \rm{DOS}^\downarrow (E_{\rm F})}{\rm{DOS}^\uparrow (E_{\rm F})+ \rm{DOS}^\downarrow (E_{\rm F})}$ = 87.1\%. The total magnetic moment is estimated to be 0.80 $\mu_B$ which deviates from the expected Slater-Pauling~\cite{galanakis2002slater,ozdougan2013slater} value of 1.0 $\mu_B$. 
Additionally, atom-specific magnetic moments were calculated, revealing values of Fe = 0.66 $\mu_B$, Ru = -0.01 $\mu_B$, V = 0.16 $\mu_B$, and Si = -0.01 $\mu_B$ for the ordered structure.

\subsection{\label{sec:XRD}X-ray diffraction}

\begin{figure}[h]
\centerline{\includegraphics[width=.48\textwidth]{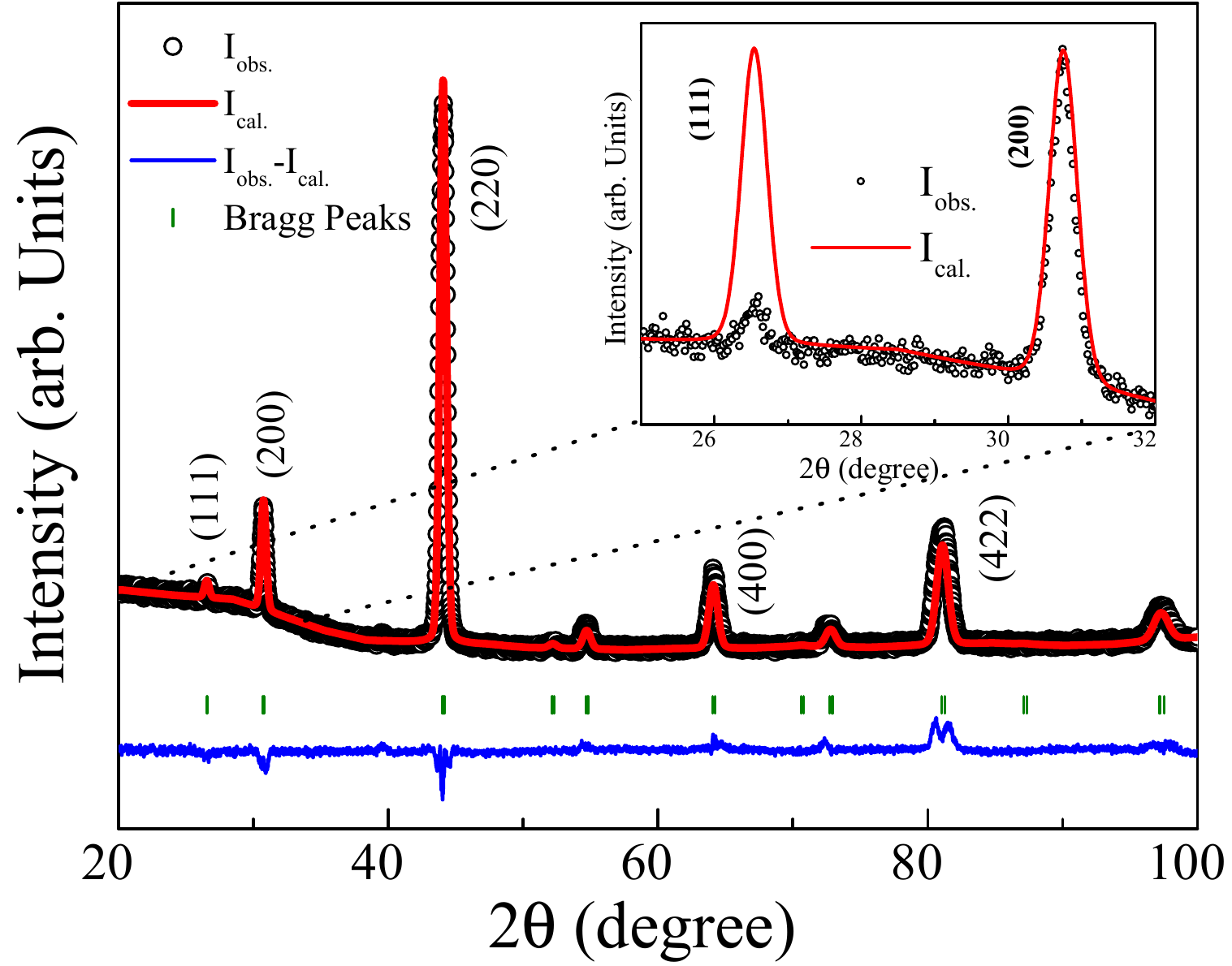}}
{\caption{Rietveld refinement of the powder XRD pattern of FeRuVSi (assuming disordered structure) measured at room temperature. Inset highlights the zoomed view of the Rietveld refinement results considering the ordered structure (Type-1).}\label{XRD}}
\end{figure}

The theoretical predictions (Table~\ref{Enthalpy}) presented above suggest that the Type-1 structure, where Si occupies the 4\textit{a} site, V occupies the 4\textit{b} site, Ru is in the 4\textit{c} site, and Fe in the 4\textit{d}, exhibits lower formation energy when compared to the other two ordered structure types. In simpler terms, the most stable ordered arrangement should consist of alternating Si/V and Ru/Fe layers, where different atoms within each layer are also arranged in a periodic manner. Such ordered atomic arrangements are expected to reflect in the experimentally obtained XRD patterns as well. However, our attempt to analyze the XRD pattern of FeRuVSi using the ordered Type-1 structure failed to properly address the reduction in the intensity of (111) Bragg peak, indicating the presence of some structural disorder in the system within the Type-1 crystal lattice arrangement. It may be pointed out here that the intensities of (111) and (200) Bragg peaks are quite sensitive to structural disorder and by studying their intensities, \textit{{vis-\`{a}-vis}} the highest intensity (220) peak, one can determine~\cite{bainsla2016equiatomic,webster1973magnetic,gupta2022coexisting} the types and nature of atomic disorders present in the material.

\begin{figure}[h]
\centerline{\includegraphics[width=.48\textwidth]{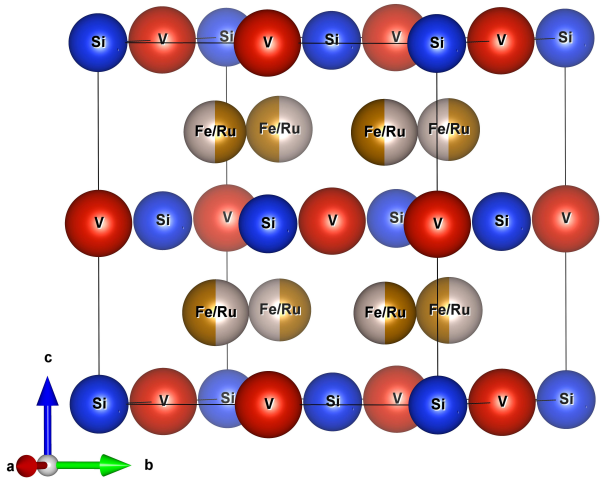}}
{\caption{Atomic arrangements of FeRuVSi in disordered structure.}\label{Disorder_Structure_Fig}}
\end{figure}

Among the various types of disorder commonly found in Heusler alloys, the A2 (space group: $Im\bar{3}m$, no. 229) and B2 (space group: $Pm\bar{3}m$, no. 221) structures are known to occur most frequently. In the A2-type structure, neither the (111) nor the (200) peaks are present, while in the B2-type structure, only the (200) peak is observed~\cite{graf2011simple,bainsla2016equiatomic,gupta2022coexisting,PhysRevB.107.184408}. In the A2-type structure, the constituent atoms ($X$, $X'$, $Y$, and $Z$) mix randomly with each other, whereas in the B2-type structure, the \textit{Y} and \textit{Z} atoms mix randomly in 4\textit{a} and 4\textit{b} sites, respectively, and the \textit{X} and $X'$ atoms mix randomly in 4\textit{c} and 4\textit{d} sites, respectively. In the XRD pattern of FeRuVSi (Fig.\ref{XRD}), both the (200) and (111) Bragg peaks are quite evident, although with slightly reduced intensity for (111) Bragg peak. Although the reduced (111) peak suggests the presence of disorder in the system, the disorder can not be either A2 or B2-type.

The reduced intensity of the (111) peak and the full intensity of the (200) peak suggest that the disorder could be of the B2-type, but of a limited nature.
We have accordingly carried out the analysis of the XRD data and found that a model in which Fe and Ru atoms mixed with each other in the 4\textit{c} and 4\textit{d} positions leads to a satisfactory fit of the pattern (Fig.~\ref{XRD}). Interestingly, this disorder between the Fe and Ru atoms transforms the ordered LiMgPdSn structure-type of quaternary Heusler alloys into a variant of the L2$_1$ structure-type, typical of the ordered ternary full Heusler alloys (Fig.~\ref{Disorder_Structure_Fig}). A similar type of disorder between the 4\textit{c} and 4\textit{d} sites atoms was previously reported in CoRuMnSi~\cite{venkateswara2020half}, FeMnVAl~\cite{gupta2022coexisting}, FeMnVGa~\cite{PhysRevB.108.045137} and NiRuMnSn~\cite{PhysRevB.108.054405}. The lattice parameter determined by Rietveld refinement is ${a=}$ 5.806(4)\,{\AA}.

If we compare the crystal structure of FeRuVSi with that of Fe$_2$VSi, we now find that in contrast to the later, the 4\textit{a} and 4\textit{b} sites in FeRuVSi have completely ordered arrangements of Si and V, respectively, whereas Fe/Ru atoms are randomly arranged in the 8\textit{c} sites of the L2$_1$ structure (equivalent to 4\textit{c} and 4\textit{d} sites in the LiMgPdSn structure-type). The strategic substitution of Fe with Ru not only solves the problem of anti-site disorder issue in Fe$_2$VSi but also demonstrates the potential of tailored disorder to modify crystal structures, providing valuable insights into the design of improved Heusler alloys.

The above-disordered structure model is further tested by theoretical estimation of the ground state energy and found to have an even lower formation enthalpy (-52.34 kJ/mol) compared to the ordered Type-1 structure (see Table~\ref{Enthalpy}), in agreement with experimental results.

\subsection{\label{sec:Mossbauer}M\"{o}ssbauer spectrometry}

\begin{figure}[]
\centerline{\includegraphics[width=.48\textwidth]{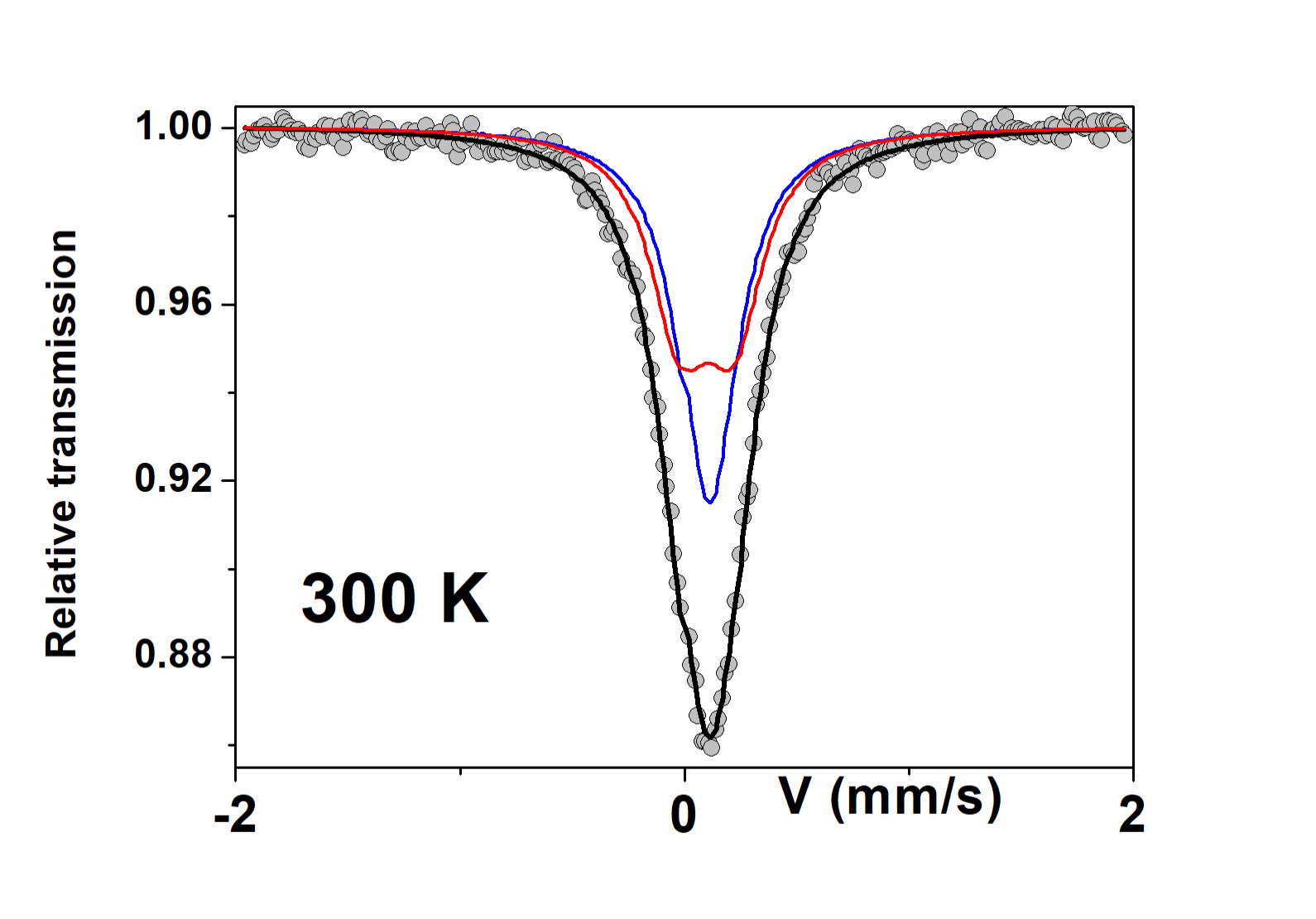}}
{\caption{M\"{o}ssbauer spectra of FeRuVSi taken at 300 K.}\label{Fig_Mossbauer}}
\end{figure}

\begin{table}[]
\caption{Fitted parameters for the M\"{o}ssbauer spectrum measured at 300 K for FeRuVSi. Isomer shift ($\delta$) (quoted relative to $\alpha$-Fe at 300 K), line-width at half height (${\Gamma}$), quadrupolar splitting ($\Delta$) and relative fractions ($\%$).}
\begin{tabular}{cccccc}
\hline
\hline
 T (K) & Site & $\delta$ (mm/s) & ${\Gamma}$ (mm/s) & {$\Delta$}                        & $\%$     \\
       &      & $\pm$0.01       &$\pm$0.01          &$\pm$0.01                         & $\pm$2\\ \hline
300    & Fe1  & 0.23            & 0.31              & 0.00                             &  50.3\%                         \\
       & Fe2  & 0.22            & 0.32              & 0.22                             &  49.7\%                         \\

\hline
\end{tabular}
\label{Mossbauer_Table}
\end{table}

To further confirm the random intermixing of Fe and Ru atoms in FeRuVSi, $^{57}$Fe M\"{o}ssbauer measurement was carried out at 300 K (Fig.~\ref{Fig_Mossbauer}). As the hyperfine structure is not well resolved, one can obtain different mathematical solutions : for the sake of simplicity in analysis, the spectrum can be characterized by two components of nearly equal intensity. Notably, even though Fe occupies a single crystallographic site in the Type-1 ordered structure, the presence of two components in the M\"{o}ssbauer spectra, with similar isomer shifts and nearly equal intensities, suggests the availability of two locations for Fe atoms with nearly identical structural environments. The spectrum suggests the presence of at least two quadrupolar doublets with small quadrupolar splitting values, as illustrated in Fig.~\ref{Fig_Mossbauer} while the values of the hyperfine parameters are given in Table~\ref{Mossbauer_Table}. The isomer shifts of these two components are almost similar, indicating similar local environments, as expected for 4\textit{c} and 4\textit{d} positions. Thus, we find that $^{57}$Fe M\"{o}ssbauer results corroborate well the structural model derived from X-ray diffraction.

\subsection{\label{sec:Magnetism}Magnetic properties}

\begin{figure}[h]
\centerline{\includegraphics[width=.48\textwidth]{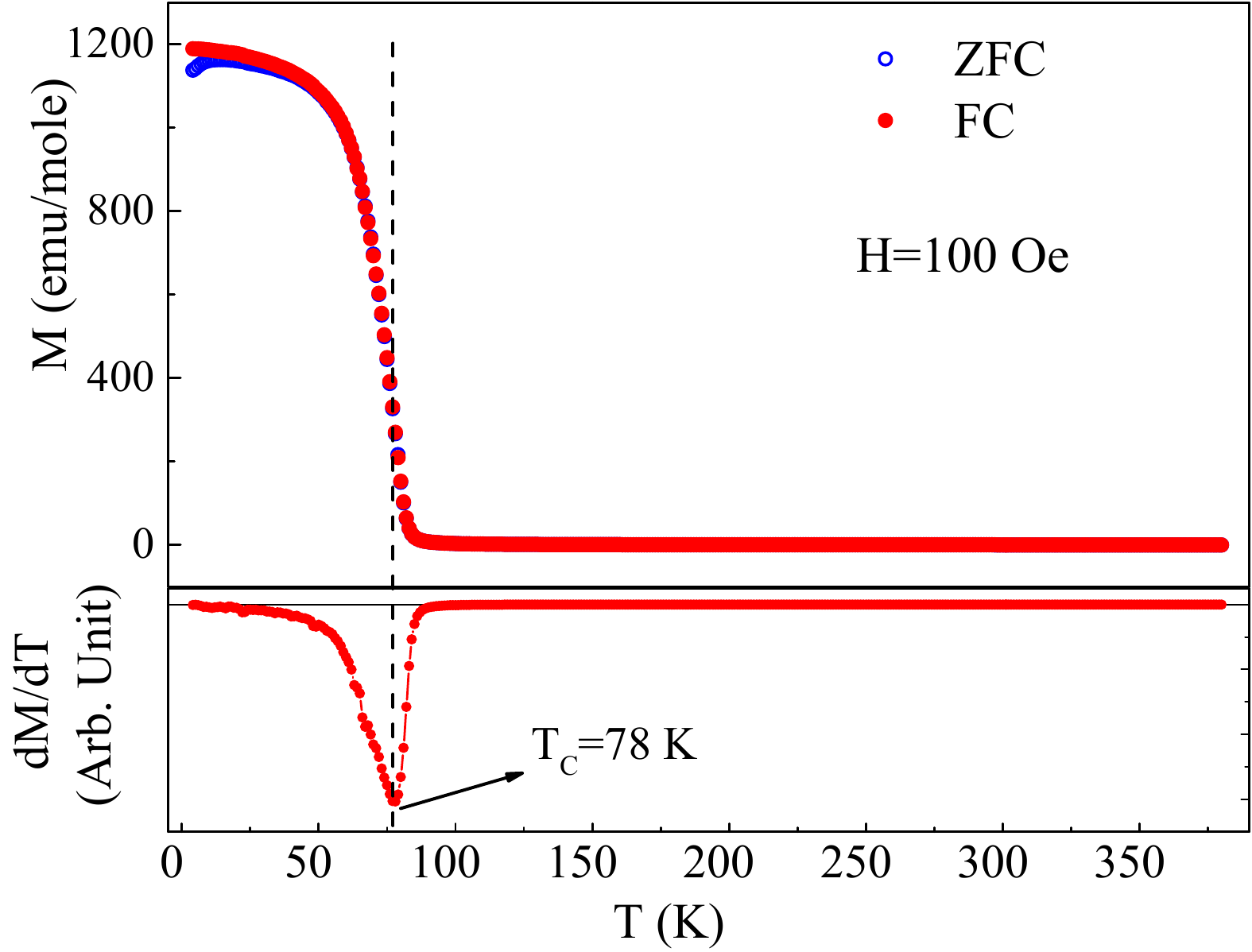}}
{\caption{Temperature dependence of magnetization in FeRuVSi measured in a 100 Oe applied magnetic field under zero field-cooled (ZFC) and field-cooled (FC) conditions. Curie temperature, T$_{\rm C}$, is determined from the minima in dM/dT \textit{vs.} T plot.}\label{MT_Fig}}
\end{figure}

\begin{figure}[h]
\centerline{\includegraphics[width=.48\textwidth]{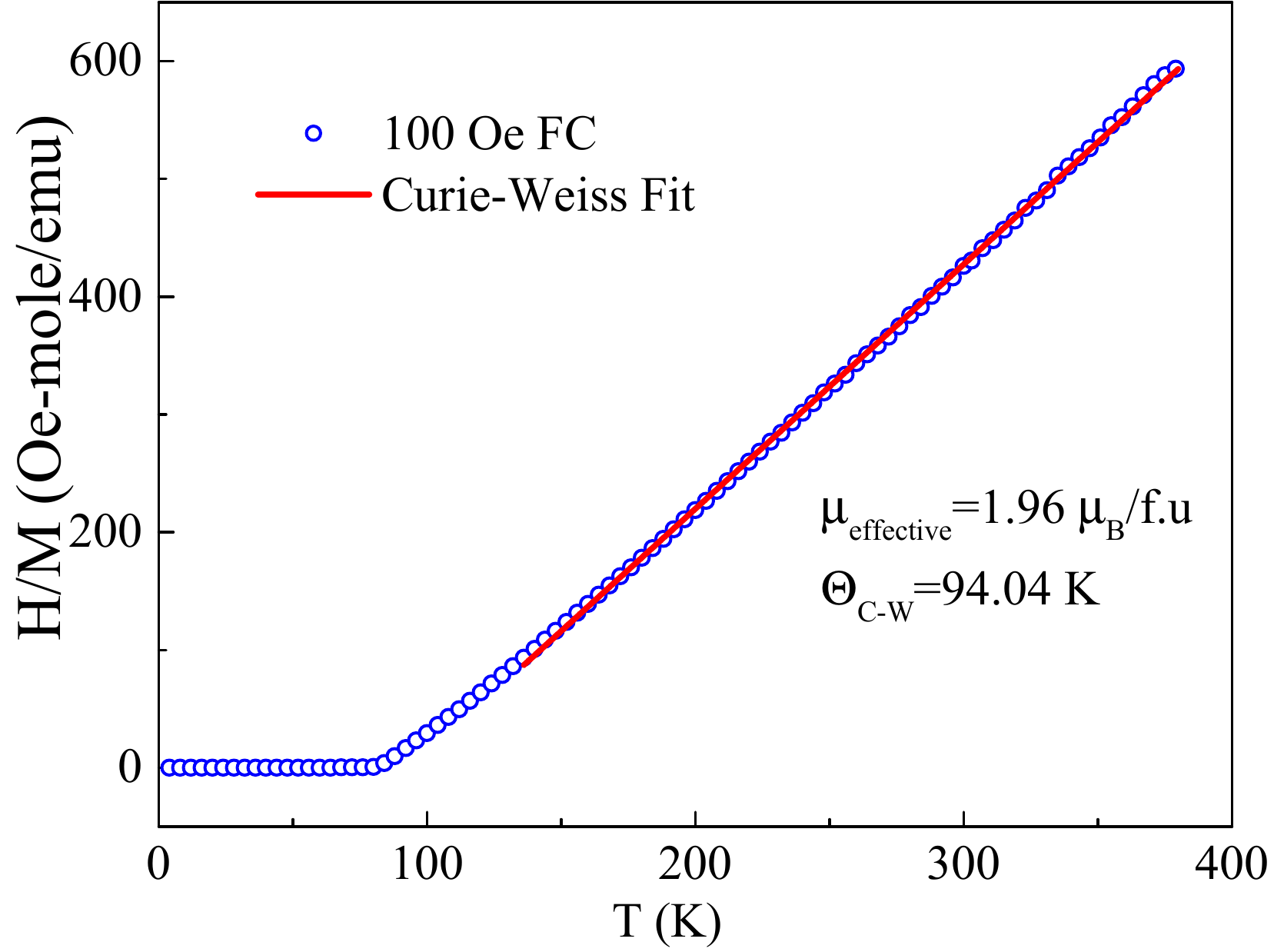}}
{\caption{Inverse susceptibility data measured under FC condition and Curie-Weiss fitting.}\label{Inverse_chi_Fig}}
\end{figure}

To determine the impact of Fe/Ru atoms distribution in 4\textit{c}/4\textit{d} Wyckoff position in magnetic properties, the temperature-dependent magnetization of FeRuVSi was measured in an applied field of 100\,Oe under both zero-field-cooled (ZFC) and field-cooled (FC) conditions (Fig.~\ref{MT_Fig}). This compound exhibits a transition from a paramagnetic (PM) to a ferromagnetic (FM) phase below a Curie temperature, denoted as $T_{\rm C} = 78$~K ($\pm{0.5}$), which we define as the temperature at which the rate of change of magnetization with respect to temperature, dM/dT, reaches its minimum.

Above the ordering temperature, the magnetic susceptibility data follows a Curie-Weiss (CW) law~\cite{gupta2022coexisting,kundu2023magnetic,chakraborty2022ground}, described by ${\chi= C/(T-\theta_P)}$, where \rm C represents the Curie constant, and $\theta_P$ is the paramagnetic Curie temperature. A linear CW fit to the inverse susceptibility (as shown in Fig.~\ref{Inverse_chi_Fig}) within the temperature range of 140-380 K yields $\theta_P$ = +94.04(3) K, with the positive sign of $\theta_P$ providing further confirmation of the presence of ferromagnetism in this material.

To determine whether the magnetism in this compound is of the localized or itinerant type, we use the Rhodes-Wohlfarth ratio (RWR). RWR is defined as the ratio of P$_C$ to P$_S$, where P$_C$ is the paramagnetic moment calculated as $\mu_{effective}^2$ = P$_C$(P$_C$+2), and P$_S$ is the saturation magnetization at low temperature. In systems characterized by localized moments, RWR tends to be close to 1, while in conventional itinerant systems, the RWR typically exceeds unity~\cite{gupta2022coexisting,PhysRevB.84.184414}. For FeRuVSi, we estimate the RWR to be 1.75, indicating that the nature of the magnetic interaction in this material is a mixed type, closer to the localized moment system.

\begin{figure}[h]
\centerline{\includegraphics[width=.48\textwidth]{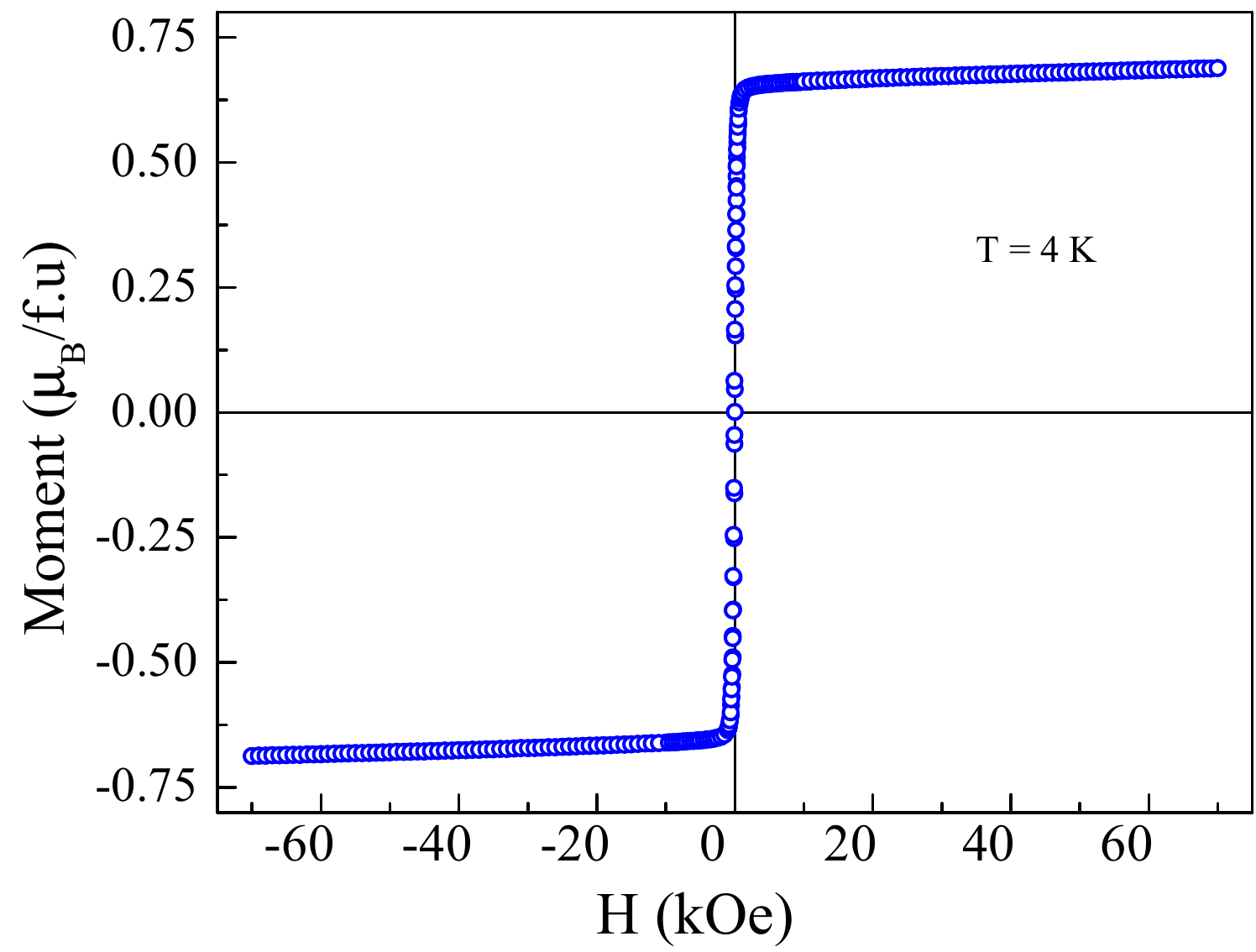}}
{\caption{Isothermal magnetization of FeRuVSi measured at 4 K\@.}\label{MH_Fig}}
\end{figure}

The Slater-Pauling rule is widely recognized as a valuable tool for elucidating the connection between a compound's magnetism (total spin-magnetic moment) and its electronic structure, particularly in the context of half-metallicity~\cite{galanakis2002slater}. Typically, it has been observed that all reported Half-Metallic Ferromagnets (HMFs) based on Heusler alloys adhere to the Slater-Pauling rule~\cite{graf2011simple,bainsla2016equiatomic}. The Slater-Pauling  rule states that the total magnetic moment in Heusler alloys can be determined by the following relationship:

\[m=(N_{V} - 24) \mu_{\rm B}/\text{f.u.},\]
\noindent
where \(N_{V}\) signifies the total count of valence electrons (VEC) for a given material~\cite{galanakis2002slater}.

In the case of FeRuVSi, the calculated VEC amounts to 25. Consequently, according to the Slater-Pauling rule, the expected magnetic moment should be 1 $\mu_{\rm B}$/f.u.. However, when scrutinizing the isothermal magnetization of FeRuVSi at 4 K (as depicted in Fig.~\ref{MH_Fig}), it becomes apparent that the material exhibits soft ferromagnetic properties with minimal hysteresis. Intriguingly, the saturation magnetization measured is 0.68 $\mu_{\rm B}$/f.u., significantly deviating from the ideal value (1 $\mu_{\rm B}$/f.u.) expected in the case of a HMF ground state for FeRuVSi.

\subsection{\label{sec:Resistivity}Resistivity}

Fig.~\ref{RT_Fig} illustrates the temperature-dependent electrical resistivity characteristics of the FeRuVSi compound in the absence of magnetic fields. The $\rho$(T) behavior demonstrates a consistent metallic nature within the measured temperature range of 5-300 K, with a distinct change in slope observed in the vicinity of the transition temperature ($T_{\rm C}$), consistent with the magnetization study.

In accordance with Mattheissen's rule, the total resistivity ($\rho (T)$) of this ferromagnetic material can be expressed as~\cite{bainsla2016equiatomic,gupta2022coexisting,PhysRevB.108.045137,gupta2023comncrga}:

\begin{figure}[h]
\centerline{\includegraphics[width=.48\textwidth]{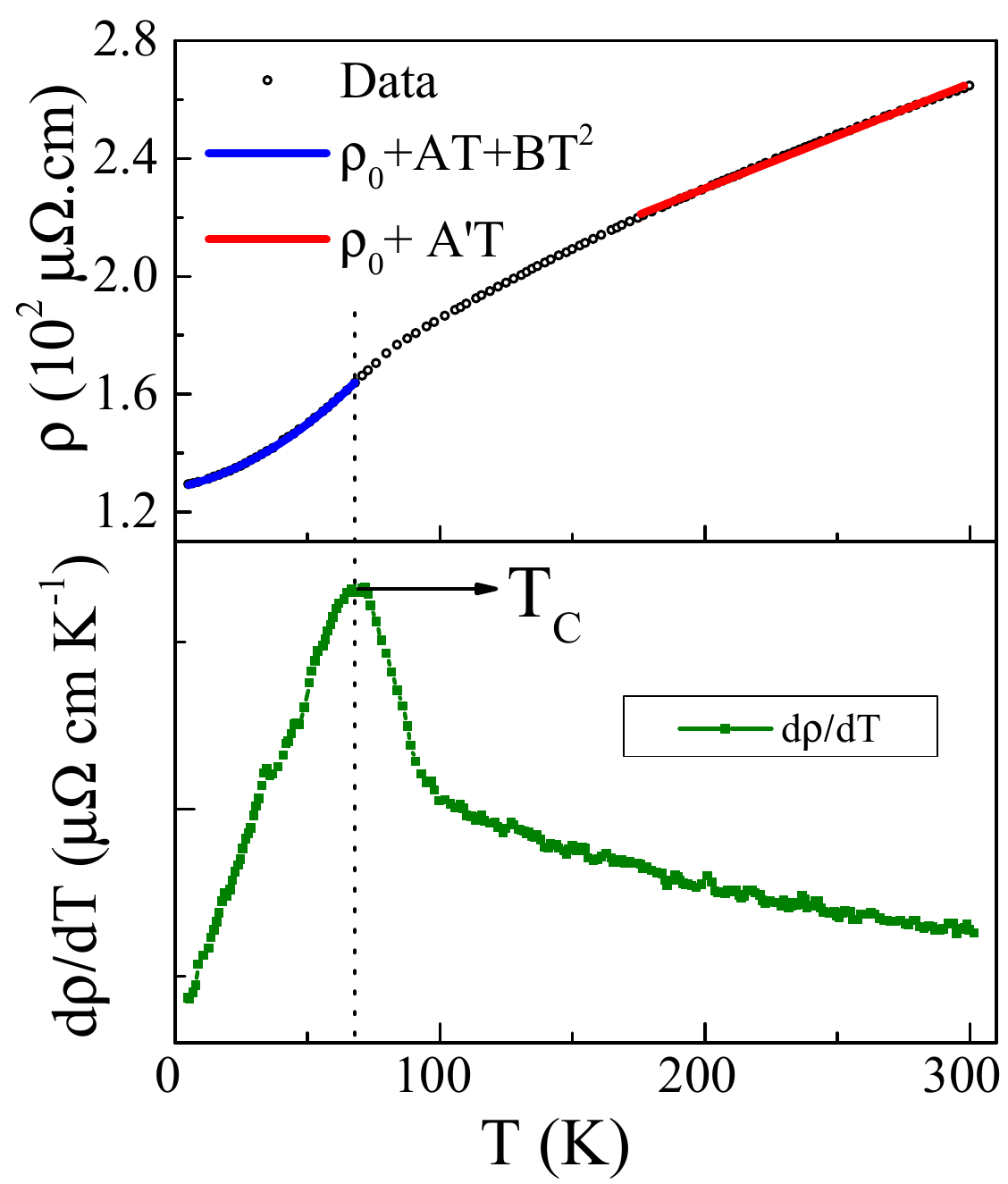}}
{\caption{Temperature dependence of electrical resistivity of FeRuVSi compound in the absence of magnetic fields. The graph exhibits metal-like resistivity behavior throughout the measured temperature region 5-300 K, with a noticeable change in slope near the transition temperature ($T_C$).}\label{RT_Fig}}
\end{figure}

\begin{equation}
\rho (T) = \rho_0 + \rho_{\text{ph}}(T) + \rho_{\text{mag}}(T)
\label{Res_Eqn1}
\end{equation}
\noindent
Here, $\rho_0$ is the residual resistivity which is temperature independent and originates from the scattering of conduction electrons by lattice defects, domain walls, dislocations, etc. {$\rho_{ph}$}(T) and ${\rho_{mag}}$(T) are the contributions from the electron-phonon~\cite{gruneisen1933abhangigkeit} and electron-magnon~\cite{bombor2013half} scattering, respectively. Depending on the region of temperature, {$\rho_{ph}$} or ${\rho_{mag}}$ dominates. In order to separate out these two contributions, the resistivity data have been analyzed separately in low and high temperature regions.

For the low-temperature data ($T < T_C$), the following eqn.~\ref{Res_Eqn2} gives a good description of resistivity:

\begin{equation}
\rho (T) = \rho_0 + AT + BT^2
\label{Res_Eqn2}
\end{equation}
\noindent
In this eqn.~\ref{Res_Eqn2}, the terms $AT$ and $BT^2$ correspond to electron-phonon and electron-magnon scattering, respectively. Notably, the $BT^2$ term is confined to temperatures up to $T_C$ due to its magnetic origin. Fig.~\ref{RT_Fig} unequivocally demonstrates that the above equation effectively fits the low-temperature data. The fitted parameters are found to be $A$=0.0018 ${\mu\Omega/K}$ and $B$= 5.12 ${\times 10^{-5}}$ ${\mu\Omega/K^{-2}}$. Furthermore, electron-phonon contributions predominate in the studied compound within this temperature range. Importantly, the presence of magnon contributions in the studied compounds is consistent with the absence of the HMF characteristics, also indicated previously by the non-adherence to the Slater-Pauling rule.

In the high temperature region ($T > T_C$), the temperature dependence of resistivity exhibits linear behavior and thus can also be explained with the following eqn.~\ref{Res_Eqn3}:
\begin{equation}
\rho (T) = \rho_0 + A'T
\label{Res_Eqn3}
\end{equation}
\noindent
where  $A'$ (= 0.0035 ${\mu\Omega/K}$) is the phonon contribution.

\subsection{\label{sec:Electronic_Structure_disordered} Effect of disorder on the electronic structure}

When examining the crystal structure of FeRuVSi, the most likely scenario involves a disordered arrangement where Fe and Ru are evenly distributed across 4$c$ and 4$d$ sites, as determined through XRD and Mössbauer spectrometry. Consequently, our electronic structure analysis has been revised to accommodate the system's disorder. Subsequent results reveal that the enthalpy of formation ($\Delta_f{H}$) for the disordered structure, as determined by SQS, is significantly lower than that of the ordered Type-1 structure, reaching -54.29\, kJ/mol . In other words, the formation energy is 2.8\, kJ/mol lower for the disordered structure compared to the ordered structure.

\begin{figure}[]
\centerline{\includegraphics[width=.48\textwidth]{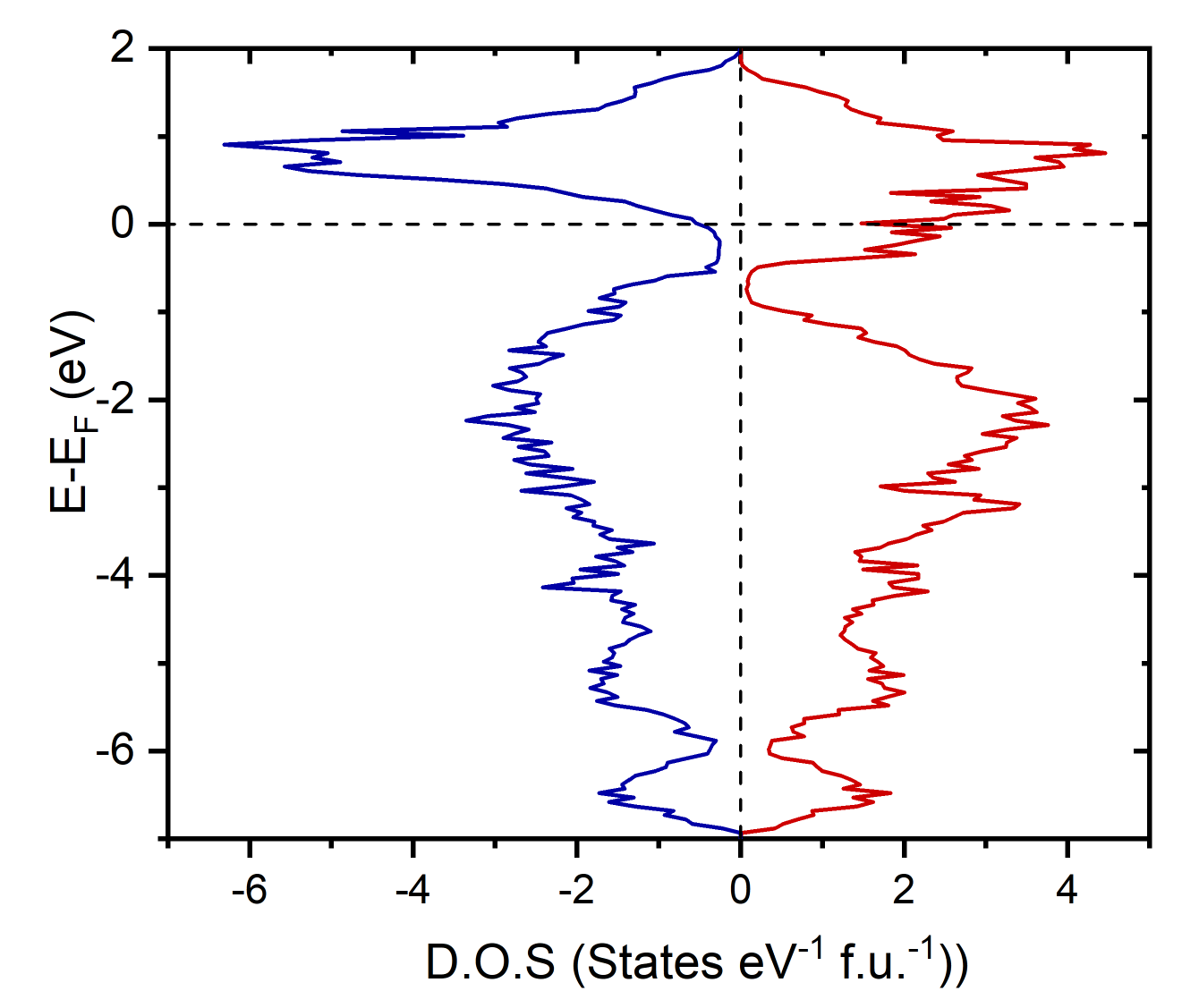}}
{\caption{Density of states of FeRuVSi for disordered structure.}\label{DOS_Disorder}}
\end{figure}

Figure~\ref{DOS_Disorder} illustrates the spin-polarized density of states (DOS) for the disordered structure. Due to the structural disorder, the spin polarization decreases to 52.9\%. As shown in Figure~\ref{DOS_Disorder}, the disordered structure maintains the ferromagnetic nature of the compound, with the total magnetic moment estimated at 0.74 $\mu_B$/f.u.. Therefore, the calculations also support the observation from our experimental data, indicating a reduction in the magnetic moment due to structural disorder.

\section{Conclusion}

Although many full Heusler alloys have been theoretically proposed to exhibit the technologically significant half-metal ferromagnetic character, actually this character is rarely achieved due to the compositional disorder inherent to these systems. A well-known example is Fe$_2$VSi, where the disorder can even transform this would-be ferromagnet into an antiferromagnet. In the present work, we have proposed a simple and novel approach to stop such atomic migration in Fe$_2$VSi by replacing half of the magnetic Fe atoms with non-magnetic Ru atoms of larger size. We have carried out theoretical DFT calculations on such a structure and found that the resulting composition is likely to exhibit ferromagnetic ordering with a large spin-polarization value. We, then synthesized FeRuVSi in polycrystalline form. Combining the XRD and $^{57}$Fe M\"{o}ssbauer measurements, we have shown that the Ru substitution successfully removes the anti-site disorder between octahedral and tetrahedral sites present in Fe$_2$VSi, as Fe/Ru atoms occupy only the tetragonal sites , albeit randomly. Magnetic measurements reveal the ferromagnetic ground state, but the isothermal magnetization at its saturation value does not reach the predicted Slater-Pauling limit. The first principle calculations, considering the actual structure model, revealed that the formation energy is such that structural disorder would be slightly more preferable to the perfectly ordered one, and is also responsible for the decrease in the spin-polarization value. Thus, the present work not only sheds light on the challenges posed by disorder in Heusler alloys but also introduces a promising approach to overcome such complexities. The innovative substitution strategy, applied here to Fe$_2$VSi, opens the doors to tailoring magnetic properties, making a significant contribution to the evolving landscape of spintronic materials.

\centerline{\textbf{Acknowledgement}}
S.G and S.C would like to sincerely acknowledge SINP, India and UGC, India, respectively, for their fellowship. DFT calculations were performed using HPC resources from GENCI-CINES (Grant 2021-A0100906175). Part of this work was performed under CSRP project 6908-3 of the Indo-French Centre for Promotion of Advanced Research, New Delhi, India.
\normalem
\bibliographystyle{apsrev4-2}
%

\end{document}